\documentclass[unsortedaddress,superscriptaddress,prb,twocolumn]{revtex4-2}
\usepackage{hyperref}
\usepackage{epsfig}
\usepackage{graphicx}
\usepackage{subfigure}
\usepackage{latexsym}
\usepackage{color}
\usepackage{fullpage}
\usepackage{dcolumn}
\usepackage{bm}
\usepackage[normalem]{ulem}
\usepackage{units}
\usepackage{amsmath}
\usepackage{amsfonts}
\usepackage[paperwidth=210mm,paperheight=297mm,centering,hmargin=2cm,vmargin=2.3cm]{geometry}

\newcommand{\johann}[1]{\textcolor{cyan}{#1}}
\newcommand{\nico}[1]{\textcolor{red}{#1}}

\begin{document}

\title{Unusual Coulomb phase physics in the arctic square ice}

\author{Johann Coraux}
\email{johann.coraux@neel.cnrs.fr}
\affiliation{Universit\'{e} Grenoble Alpes, CNRS, Institut NEEL, Grenoble INP, 38000 Grenoble, France}
\author{Nicolas Rougemaille}
\affiliation{Universit\'{e} Grenoble Alpes, CNRS, Institut NEEL, Grenoble INP, 38000 Grenoble, France}

\begin{abstract}
The square ice is a two-dimensional spin liquid hosting a Coulomb phase physics.
When constrained under specific boundary conditions, the so-called domain wall boundary conditions, a phase separation occurs which leads to the formation of a spin liquid confined within a disk surrounded by magnetically ordered regions.
Here, we numerically characterize the ground state properties of this spin liquid, coined the arctic square ice in reference to a phenomenon known in statistical mechanics.
Our results reveal that both the vertex distributions and the magnetic correlations are inhomogeneous within the liquid region, and exhibit a radial dependence. 
If these properties resemble those of the conventional square ice close to the center of the disk, they evolve continuously as the disk perimeter is approached.
There, the spin liquid orders.
As a result, pinch points, signaling the presence of algebraic spin correlations, coexist with magnetic Bragg peaks in the magnetic structure factor computed within the disk.
The arctic square ice thus appears as an unconventional Coulomb phase sharing common features with a fragmented spin liquid, albeit on a charge neutral vacuum.
\end{abstract}

\maketitle

\section{Introduction} 
\begin{figure}[!hbt]
\begin{center}
\includegraphics[width=8cm]{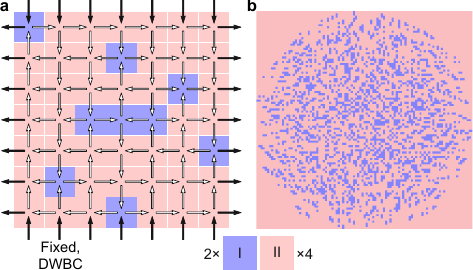}
\caption{\label{fig1} (a) Schematics of the square ice model under domain wall boundary conditions. The spins at the lattice edges (in black) are fixed, whatever the temperature. Ice rule vertices correspond to four spins arranged in such a way that their divergence is zero at each vertex site (where two spins point inwards and two spins point outwards). Numbers indicate the degeneracy of each vertex type. (b) A randomly chosen ground state configuration of the square ice model under DWBC. The lattice contains $101 \times 101$ vertices, i.e., 20604 Ising spins.}
\end{center}
\end{figure}

When studying the thermodynamic properties of a spin model, one generally seeks to limit finite size effects, i.e., the impact of boundary conditions.
On the contrary, certain models exhibit fascinating behaviors specifically when peculiar boundary conditions are chosen.
This is the case of the square ice model, known in statistical mechanics to present a zero-point entropy \cite{Pauling1935, Nagle1966, Lieb1967a, Lieb1967b} and in frustrated magnetism to be the archetype of a two-dimensional Coulomb phase \cite{Perrin2016, Ostman2018, Farhan2019, Rougemaille2019, Perrin2019, Rougemaille2021, Brunn2021, Goryca2021, Schanilec2022}.
The square ice is a classical spin liquid whose magnetic correlations decrease algebraically with the distance, and in which excitations behave as magnetic monopoles \cite{Ryzhkin2005, Castelnovo2008, Henley2010, Moessner2016}.
When this spin model is subjected to boundary conditions for which the spins at the edges of the lattice are fixed, whatever the temperature, in the so-called domain wall boundary conditions (DWBC) [see Fig.~\ref{fig1}(a)], a phase separation mechanism occurs \cite{King2023}.
In its ground state, the center of the lattice is massively disordered, greatly fluctuates, and is composed of an almost random (yet constrained) arrangement of ice-rule obeying vertices [the type I and type II vertices represented in Fig.~\ref{fig1}(a) in blue and red, respectively, which satisfy a local divergence-free condition of the spin arrangement].
On the contrary, the outermost regions of the lattice are frozen, magnetically ordered and composed only of type II vertices.

What is remarkable in this phase separation is that the ordered regions are not just confined along the lattice edges where the spins are fixed, but rather extend into the system in such a way that the interface separating the ordered and disordered regions is relatively well described by a circle inscribed inside the lattice.
This astonishing property can be visualized directly by randomly selecting one ground state configuration in a fairly large lattice, simply representing the spatial distribution of the two possible vertex types [see Fig.~\ref{fig1}(b)].
What is also remarkable is that its properties do not depend on the lattice size and persists at the thermodynamic limit.
The square ice under DWBC is characterized by a zero-point entropy per spin $\mathcal{S}^{DWBC}=\frac{1}{4}ln(\frac{3^3}{2^4})\sim 0.131$ \cite{Tavares2015} smaller than that of the conventional square is ($\mathcal{S}=\frac{3}{4}ln(\frac{4}{3})\sim 0.216$, see Ref. \onlinecite{Lieb1967a}) due to the partial crystallization of the lattice.

\begin{figure*}
\begin{center}
\includegraphics[width=10cm]{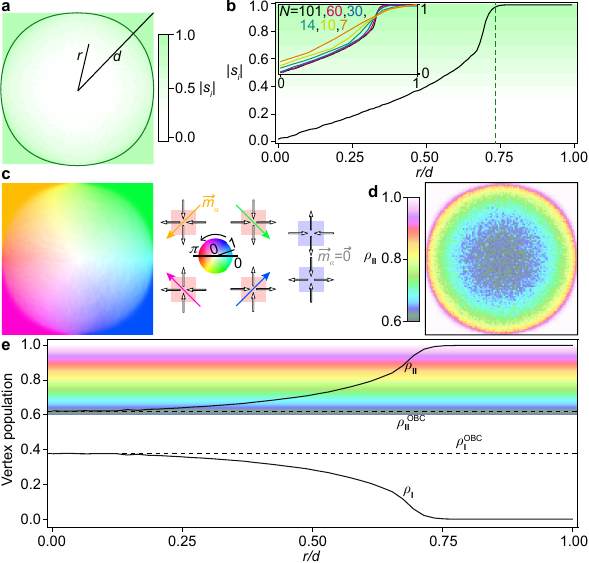}
\caption{\label{fig2} (a) Contour plot of the average modulus $|\vec{s}_i|$ of the spin located on site $i$ ($i\in [0,2N(N+1)]$) in a lattice containing $N \times N$ vertices ($N=101$). The solid line on the map represents the arctic circle analytically known for infinite-size lattices. (b) Intensity cut along the lattice's half diagonal and deduced from (a); Inset: same data for several $N$ values. As $N$ increases, the transition from the disordered to ordered regions sharpens. The dashed line shows the position of the interface between the disordered and ordered phases. (c) Color-wheel map showing the orientation and amplitude of the magnetic moment carried by the type II vertices. (d) Corresponding radial distributions of type II vertices [$\rho_\mathrm{II}(r/d)$ with $d$ the length of the lattice's half diagonal]. (e) Intensity cut along the lattice's half diagonal and deduced from (d). The dashed lines indicate the values of the $\rho_\mathrm{I}$ and $\rho_\mathrm{II}$ populations expected in the square ice under open boundary conditions (OBC).}
\end{center}
\end{figure*}

\bigskip

Although first discovered in combinatorial mathematics \cite{Propp1992, Propp, Johansson2002, Zinn2010} and subsequently studied in statistical physics \cite{Korepin2000, Kenyon2006, Cugliandolo2015, Cugliandolo2017}, this behavior, coined the arctic circle phenomenon, presents unique properties when approached under the prism of a frustrated spin model.
In particular, a spin model de facto introduces an energy scale, absent in domino or vertex tiling problems, and thus allows to investigate the properties of magnetic excitations in a phase separated system.
A programmable lattice of superconducting qubits \cite{King2023} recently provided a first experimental implementation of this physics, revealing two main results:

\noindent - The spin liquid in the interior of the arctic curve differs from the conventional square ice as the constraint imposed by the DWBC extend deep into the lattice.
As a consequence, the spin liquid is dressed with an average spin texture which reflects the spin direction fixed at the lattice edges.
We will describe this average spin texture in more details in Sec. III.

\noindent - At low temperature, magnetic monopoles were found to accumulate along the four edges of the lattice, but certain edges are preferred depending on the magnetic charge and magnetic moment the monopoles carry. 
Strikingly, this charge and moment selection occurs without the application of any external driving force, but is rather inherited from the average spin texture dressing the spin liquid manifold.

In that work however, relatively small size lattices were studied, at very low but non-zero temperature, i.e., with a vanishing but non-zero density of monopole excitations.
Here, we pursue this initial study and investigate numerically the \textit{ground state} properties of the arctic square ice using a loop flip algorithm to shuffle the spin lattice.
In other words, we disregard any energy scale and magnetic excitations are not considered.
The spin liquid properties are then characterized through the computation of the local magnetization, the vertex populations and pairwise spin correlations.
Our analysis shows that all these quantities radially evolve in the interior of the arctic curve and differ from those of the square ice.
Besides, the magnetic structure factor calculated within the spin liquid region reveals the coexistence of spin order and spin disorder, evidenced by the presence of magnetic Bragg peaks and a diffuse, yet structured intensity.
Similar to the square ice, pinch points are observed at certain locations in reciprocal space, revealing the algebraic nature of the magnetic correlations.
The arctic square ice thus shares common features with fragmented spin liquids, i.e., spin liquids characterized by an ordered magnetic component coexisting with a highly fluctuating Coulomb manifold.
Nevertheless, as mentioned above, magnetically charged excitations are not considered in our shuffling algorithm and the ice rule contraint is strictly obeyed. 
Magnetic ordering in the arctic square ice is not associated to a divergence full channel resulting from an Helmholtz decomposition of the spin vector field \cite{Brooks2014}, but rather originates from the sole boundary conditions.

\begin{figure}
\begin{center}
\includegraphics[width=8cm]{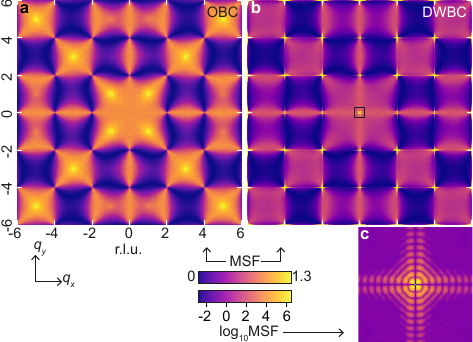}
\caption{\label{fig3} Magnetic structure factors (MSFs) of the square ice (a) and arctic square ice (b,c). The MSFs are calculated for a lattice containing $101 \times 101$ vertices. The color scale is identical in (a,b) and illustrates that the intensity within the diffuse background is substantially lower in the arctic square ice. The map in (b) also reveals Bragg-like features at the zone center ($[q_x,q_y]=[2r,2s]$ wavevectors, with $(r,s)\in\mathbb{Z}^{*2}$), as expected from the magnetically ordered regions surrounding the arctic curve. A zoom highlighting the fine structure \nico{[black square in (b)]} of these features in shown in (c).}
\end{center}
\end{figure}

\section{Numerical approach}

The numerical approach we followed consists in generating a set of $10^3$ ice rule satisfying configurations using a loop flip algorithm, i.e., an algorithm in which the update consists in reversing the direction of all the spins contained in an oriented closed loop.
Contrary to a single spin flip dynamics, flipping such a loop always preserves the ice rule contraint, and thus allows us to probe the ground state manifold.
In practice, $n$ loop updates are performed for each stored configuration, starting from a randomly chosen one complying with the DWBC.
Typically, $n = N^2$ loops are required to shuffle efficiently a $N \times N$ vertex lattice [i.e., containing $2N(N+1)$ spins] under open boundary conditions \cite{percolation}.
Under DWBC, the shuffling process is less efficient because closed loops become statistically rare in the four lattice corners, where a fully polarized state develops, and larger $n$ values are required to shuffle the configurations.
Most results reported in this work were obtained for $N=101$ and $n$ was set to $3 \times 10^6$.
Smaller lattice sizes were also studied [Fig.~\ref{fig2}(b)].

We emphasize again that there is no energy scale in our approach and monopole excitations are not considered.
Instead, our algorithm allows to randomly select a set of ground state configurations.
As a consequence, only type I and type II vertices are present in the selected configurations.
The set of ice rule-obeying configurations is finally used to compute the average local magnetization, the vertex populations and the magnetic structure factor.

\begin{figure*}
\begin{center}
\includegraphics[width=132.7mm]{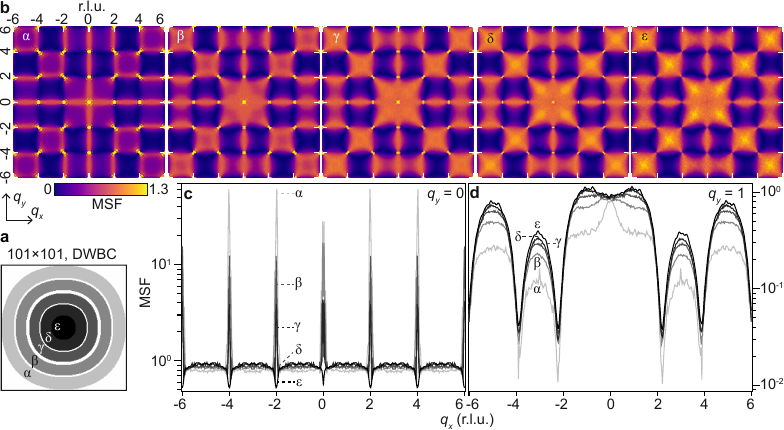}
\caption{\label{fig4} (a) Schematics showing five annular regions in which the MSF is calculated for a lattice containing $101 \times 101$ vertices. (b) MSFs computed in the five annular regions shown in (a) and revealing how magnetic correlations evolve from the lattice center to the arctic curve. Close to the lattice center, the MSF strongly resemble the one of the square ice. However, magnetic Bragg peaks at the center of the Brillouin zone pop up as the arctic curve is approached, at the expense of antiferromagnetic correlations that substantially decrease at the corners of the Brillouin zone. The ripples that can be observed around the magnetic Bragg peaks originate from the form factor, i.e., are a consequence of computing the Fourier transform within an annular region. (c,d) Intensity cuts along peculiar directions in reciprocal space showing quantitatively how the intensity changes in the MSFs reported in (b).}
\end{center}
\end{figure*}

\section{Magnetic properties of the arctic square ice}
\subsection{Average spin texture}

Analyzing our set of $10^3$ ground state configurations, we first calculate the average spin value on each lattice site and generate a two-dimensional colored map of the local spin magnitude $|\vec{s}|$ [see Fig.~\ref{fig2}(a)].
In such a map, $|\vec{s}|$ varies between 0 (white) and 1 (light green), 1 meaning that the spin direction does not change in all shuffled configurations (as we expect at the lattice edges due to the boundary conditions), while 0 reflects perfect balance between the two possible spin directions.
As anticipated, a phase separation is evidenced: the center of the lattice fluctuates greatly ($|\vec{s}|$ is small), whereas the four lattice corners are frozen ($|\vec{s}|=1$).
Moreover, the interface delimiting the frozen and fluctuating regions resembles an inscribed circle [see black contour in Fig.~\ref{fig2}(a)].
It is worth mentioning that the curve calculated at the thermodynamic limit ($N\rightarrow+\infty$) is not a circle, but rather consists of four portions of an ellipse \cite{Cugliandolo2015}.
We also stress that $|\vec{s}|$ varies gradually from 0 to 1, as shown by the evolution of the green shade contrast in Fig.~\ref{fig2}(a).
This is better illustrated by plotting the average $|\vec{s}|$ radial distribution [see Fig.~\ref{fig2}(b)].
The curve unambiguously reveals the phase separation.
$|\vec{s}|$ is in fact equal to 0 only at the lattice center and increases almost linearly with the distance $r$.
For $r/d \sim 1/\sqrt2$, $d$ being half of the lattice diagonal, $|\vec{s}|$ increases abruptly to reach 1, illustrating that the constraint imposed by the DWBC propagates substantially into the lattice.

It is also instructive to represent the average local magnetization $\vec{m}$ that can be assigned to each vertex.
Indeed, if type I vertices carry no net magnetic moment because of their antiferromagnetic configurations, type II vertices do.
The color-wheel map in Fig.~\ref{fig2}(c) reveals two interesting features.
First, the four corners of the lattice are each populated by one of the four possible type II vertices, the one matching the local constraint imposed by the boundary conditions.
Each corner is thus magnetically ordered but differs from the other three \cite{King2023}.
It is interesting to note that the magnetic configurations within the four corners strongly influence the magnetization texture in the interior of the arctic curve, and the whole lattice can be divided into four quadrants statistically favoring one of the four type II vertices [see how the colors are distributed in Fig.~\ref{fig2}(c)].
Second, the map reveals that $\vec{m}$ winds around the lattice, turning by a $2\pi$ angle on a closed path with a winding number $-1$. 
This spin texture is inherited from the DWBC, where the orientation of $\vec{m}$ already turns by $2\pi$. 
Overall, the spin texture resembles an anti-vortex \cite{King2023}.

\subsection{Vertex populations}

As revealed by the contour plot [Fig.~\ref{fig2}(a)] and the color-wheel map [Fig.~\ref{fig2}(c)], the contraint imposed by the lattice boundaries extends deep into the system.
In this section, we investigate how the DWBC impact the vertex populations in the interior of the arctic curve and compute the radial density of type I and type II vertices.
The results are reported in Fig.~\ref{fig2}(d) and reveal an interesting property: the vertex populations vary gradually from the lattice center to the lattice edges, and three regimes may be defined:

\noindent - When $r/d<1/5$, the vertex populations are almost constant and hardly distinguishable from those of the conventional square ice [the fractions of type I ($\rho_\mathrm{I} $) and type II ($\rho_\mathrm{II}$) vertices are close to the 38$\%$ / 62$\%$ fractions expected in the square ice].
In fact, careful inspection of the curves shows that the vertex populations slightly vary, but the change is small.

\noindent - When $r/d>3/4$, the vertex populations are also constant, $\rho_\mathrm{I}=0$ and $\rho_\mathrm{II}=1$: the spin texture is locally ordered and consists of a type II vertex tiling.

\noindent - When $1/5<r/d<3/4$, the vertex populations show a continuous change between what is expected in the square ice and the ordered configuration.
We note that if they slowly vary for low $r/d$ values, their change is more abrupt when approaching the arctic curve ($r/d \sim 1/\sqrt2$).

Close to the lattice center, we then expect to find the properties of the square ice.
However, discrepancies with the conventional Coulomb phase for $1/5<r/d<3/4$ are expected, and are likely more pronounced as the arctic curve is approached.

\begin{figure*}
\begin{center}
\includegraphics[width=132.6mm]{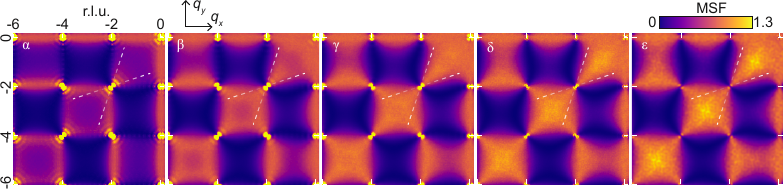}
\caption{\label{fig5} Magnetic structure factor computed for the five annular regions reported in  Fig.~\ref{fig4}. Although the pinch points cannot be resolved due to the presence of the magnetic Bragg peak, their width, indicated by a white dashed line, does not change in the five annular regions.}
\end{center}
\end{figure*}

\subsection{Magnetic structure factor}

To further analyze the properties of the spin liquid in the interior of the arctic curve, we now investigate the spin correlations through the computation of the magnetic structure factor (MSF).
The MSF is defined in the sense of neutron scattering experiments, in which magnetic correlations \textit{perpendicular} to the scattering vector $\vec{q}$ are measured. 
Introducing the perpendicular spin component $\vec{S}^\perp$
\begin{equation}
\vec{S}^\perp=\vec{S}-(\vec{\hat{q}}\cdot\vec{S})\vec{\hat{q}}
\label{sip}
\end{equation}
where $\vec{\hat{q}}$ is the unit vector along $\vec{q}$, the scattered intensity can be written as:
\begin{equation}
I(\vec{q})=\frac{1}{\mathcal{N}} \sum_{i,j} \vec{S}_{i}^\perp\cdot\vec{S}_{j}^\perp \exp^{i\vec{q}\johann{\cdot}\vec{r}_{ij}}
\label{I}
\end{equation}
where $\mathcal{N}$ is the total number of spins and $\vec{r}_{ij}=\vec{r}_i - \vec{r}_j$.

As a reference, we have first calculated the MSF under open boundary conditions to obtain the structure factor of the square ice [Fig.~\ref{fig3}(a)] \cite{note_rlu}. 
Consistent with previous works, this MSF exhibits the well-known diffuse structured pattern \cite{Perrin2016, Rougemaille2021} with pinch points signaling the presence of algebraic magnetic correlations in real space [in Fig.~\ref{fig3}(a) see $[q_x,q_y]=[2r,2s]$ wavevectors, with $(r,s)\in\mathbb{Z}^{*2}$].

Now computing the MSF for the square ice under DWBC, a different pattern is observed [Fig.~\ref{fig3}(b)].
Although the MSF still exhibits a diffuse and structured pattern, it additionally features magnetic Bragg peaks at the $\Gamma$ point ($[q_x,q_y]=[2r,2s]$ wavevectors, with $(r,s)\in\mathbb{Z}^{2}$) and fainter intensity at the corners of the Brillouin zone ($[q_x,q_y]=[2r+1,2s+1]$ wavevectors, with $(r,s)\in\mathbb{Z}^{2}$).
This result is consistent with the vertex population analysis provided in Sec.~III.B, which illustrates the statistical importance of the type II ordered regions surrounding the arctic curve.

Based on the conclusion deduced from the vertex analysis, the MSF is recomputed in the interior of the arctic curve, but within a ring of width $(r_1-r_2)/d$, as schematized in [Fig.~\ref{fig4}(a)].
This allows us to characterize the magnetic correlations within annular regions in which the vertex populations do not vary in their entire range, while keeping a statistics sufficiently large.
The results are reported in Figs.~\ref{fig4}(b) for five rings corresponding to $(r_1/R,r_2/R)$ = (1,0.8), (0.77,0.61), (0.56,0.40), (0.37,0.20), (0.20,0), with $R$ the arctic curve's radius.

First, we observe that all MSFs differ.
Consistent with the vertex population analysis [see Fig.~\ref{fig2}(d)], they reveal that magnetic correlations are not homogeneous in the interior of the arctic curve but rather exhibit a radial dependence.
Close to the lattice center, magnetic correlations are almost indistinguishable from those of the conventional square ice [compare Figs.~\ref{fig3}(a) and ~\ref{fig4}(b)].
However, as we gradually move away from the lattice center, antiferromagnetic correlations, visible at the corners of the Brillouin zone and associated to the presence of type I vertices, decrease at the expense of ferromagnetic correlations [for example, see how intensity at $[q_x,q_y]=[1,1]$ varies between the five cases shown in Figs.~\ref{fig4}(b)].
Comparison with the conventional square ice can be made more quantitative by plotting intensity cuts along specific reciprocal space directions, especially along $[q_x,0]$ and $[q_x,1]$ [see Figs.~\ref{fig4}(c) and \ref{fig4}(d), respectively].

\subsection{Pinch point analysis}

We note that the ferromagnetic correlations show up in reciprocal space precisely where pinch points signal the presence of a Coulomb phase in the square ice.
We might then wonder whether pinch points can still be observed in the arctic square ice, despite the emergence of Bragg peaks. 
We thus have recomputed the MSF around the $[q_x,q_y]=[2,2]$ point, for the square ice and for the five annular regions [see Fig.~\ref{fig5}].
Although pinch point singularity cannot be resolved, no broadening of its width can be evidenced compared to the square ice case (see the white contours highlighted in Fig.~\ref{fig5}).
This result suggests that the spin liquid in the interior of the arctic curve could be a Coulomb phase.

We also note that the magnetic Bragg peaks emerging at the centers of the Brillouin zone actually consist of two sub-peaks (except at the $\Gamma$ point where the Bragg peak intensity is composed of four sub-peaks, Fig.~\ref{fig3}(c)). 
Similar effects where observed experimentally in artificial kagomé spin ice systems \cite{Schanilec2020, Yue2022, Hofhuis2022}.
Using the language of diffraction, this fine structure within the Bragg peaks can be ascribed to interferences in the scattering from different underlying orders that are not homogeneously distributed in space. 
These underlying orders have, in each of the four quadrants within the arctic curve, a tendency to favor prominently one of the four type II vertices.

\begin{figure*}[!hbt]
\begin{center}
\includegraphics[width=111mm]{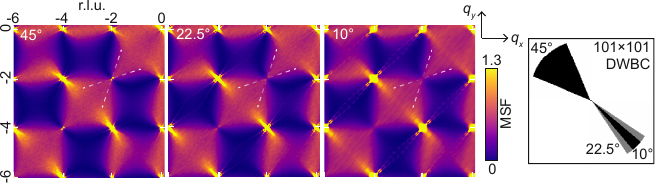}
\caption{\label{fig6} Magnetic structure factors computed for conical sectors within the arctic circle, chosen to resolve the intensity around pinch points. The axis of the conical sectors makes a $45^\circ$ angle with respect to the main directions of the square lattice, and the sectors cover different angular zones (cone angle of {10$^\circ$, 22.5$^\circ$, and 45$^\circ$, see right cartoon}). Noticeably, owing to the cone direction, the MSFs behave differently along the [1 1] and [-1 1] orthogonal directions. In the [1 1] direction, pinch points are unambiguously revealed (see white dashed line), whereas they are buried under intense magnetic Bragg peaks in the [-1 1] direction. Similar to Fig.~\ref{fig4}, the shape of the magnetic Bragg peaks originates from the form factor (Fourier transform) of the conical sectors.}
\end{center}
\end{figure*}

At this point, it is instructive to remind that the magnetic structure factor, as defined here \nico{(Eq.~\ref{I})}, probes the spin correlations perpendicular to the wavevector $\vec{q}$.
Since the spin texture in the interior of the arctic curve is anisotropic, we can compute it in such a way that $\vec{q}$ is orthogonal to the average spin direction.
This can be done by calculating the MSF within a sector making an angle $\alpha$ with the average spin direction, as represented in Fig.~\ref{fig6}.
The MSF is then also anisotropic, showing different behavior  in the [1 1] and [-1 1] directions.
Indeed, if the diffuse intensity does not change, the magnetic Bragg peaks are then less visible in one of the two orthogonal directions.
This asymmetry in the Bragg peak intensity is more pronounced as $\alpha$ is reduced.
Although the statistics decreases with $\alpha$, the MSF unambiguously reveals the presence of pinch points buried under the Bragg peaks, elegantly demonstrating the algebraic nature of the magnetic correlations in the arctic square ice.

\section{Summary and discussion}

In this work, we used a loop flip algorithm to explore the ground state manifold of the square ice model under domain wall boundary conditions.
Under such conditions, a phase separation occurs in which a spin liquid is confined in the interior of an inscribed circle and surrounded by four magnetically ordered regions.
Analyzing the local magnetization, the vertex populations and magnetic correlations within the spin liquid, we find that all quantities exhibit a radial dependence. 
The properties of the confined region strongly resemble those of the conventional square ice close to the lattice center, but magnetic ordering gradually develops as the lattice edges are approached.
Despite these strong variations, pinch points are evidenced in the magnetic structure factor, suggesting that the arctic square ice is a Coulomb phase, albeit different from the one known for the conventional square ice.
Algebraic spin-spin correlations thus coexist with magnetic Bragg peaks within the arctic square ice.

In other words, the arctic square ice is at the same time ordered and disordered, and spins strongly fluctuate on top of an average spin texture that reflects the constraint imposed by the boundary conditions.
In that sense, the arctic square ice presents similarities with a fragmented spin liquid \cite{Brooks2014, Canals2016, Petit2016, Lefrancois2017, Elsa2020}.
However, in a fragmented spin liquid the ordered spin component is homogeneously distributed in space and corresponds to an alternation of all-in / all-out configurations at each vertex sites.
The ordered spin component is a crystal of opposite magnetic charges \cite{Rougemaille2011, Zhang2013, Chioar2014, Montaigne2014, Drisko2015}.
In the arctic square ice, the ordered spin component varies continuously and radially from the lattice center, where it vanishes completely, to the lattice edges, where it saturates.
Nevertheless, contrary to the fragmented square ice, here the ordered component only results from a local divergence free condition and not from the injection of magnetic charges.
Order and disorder thus coexist within a charge neutral vacuum and do not show up after an Helmholtz decomposition.

It is interesting to note that the presence of a Coulomb phase physics in the square ice under DWBC may have been anticipated.
Indeed, this is what should be observed for a vector field characterized by a local divergence-free constraint and describing an extensively degenerate manifold.
However, we should keep in mind that these conditions alone are not sufficient, and the disordered manifold \textit{must} behave like a paramagnet.
In other words, corse-grained regions of the lattice must be statistically independent \cite{Henley2010}.
This is not the case under DWBC as magnetic order develops when approaching the lattice edges (i.e., neighboring coarse-grained regions are strongly correlated one another).
Since these ordered correlations cannot be separated from the disordered component through an Helmholtz decomposition, Henley's rules are not fulfilled and there is not reason a priori to expect a Coulomb phase physics in the arctic square ice.

Several questions may now come to mind that could stimulate further theoretical and experimental investigations. 
For example, we might wonder to what extent an external applied magnetic field would affect the phase separation mechanism and the spatial extension of the ordered and disordered phases. 
In particular, understanding whether an applied field could counterbalance the formation of an arctic curve would definitively be an interesting route to pursue in the future.

\section{acknowledgements}
This work was supported by the Agence Nationale de la Recherche through Projects No. ANR-22-CE30-0041-01 “ArtMat.” 
The authors thank Andrew King and Benjamin Canals for fruitful discussions.


\begin{thebibliography}{2}
\bibitem{note_rlu}
The square lattice consists of two sub-lattices, one bonding the vertical spins, the other bonding the horizontal ones. Two neighboring spins from the same sub-lattice are separated by a distance $a$. The unit cell we consider has $a2$ area and comprises one vertical spin and one horizontal spin. The principal axes of the unit cell are aligned with those of the square lattice. The corresponding Brillouin zone thus extends in the $[-\pi /a , +\pi /a]$ interval in the two perpendicular directions, $\pi /a$ defining the reciprocal lattice unit (r.l.u).
\end{thebibliography}
\end{document}